\input{aipcheck}

\documentclass[
    ,final            
  ]
  {aipproc}

\layoutstyle{8x11single}

\begin{document}

\title{Hard Probes of the Quark Gluon Plasma in Heavy Ion Collisions}

\classification{<Replace this text with PACS numbers; choose from this list:
                \texttt{http://www.aip..org/pacs/index.html}>}

\author{Carlos A. Salgado}{
  address={Departamento de F\'\i sica de Part\'\i culas and IGFAE, Universidade de Santiago de Compostela\\ E-15782 Santiago de Compostela (Galicia-Spain)}
}

\begin{abstract}
The medium-modifications of processes characterized by the presence of a hard scale provide the most diverse tools to characterize the properties of the  matter created in high-energy nuclear collisions. Indeed, jet quenching, the suppression of particles produced at high transverse momentum, has been established at RHIC almost a decade ago as one of the main tools in heavy-ion collisions. The melting of quarkonia is expected to provide also information about the temperature and the properties of the produced medium. The beginning of the LHC era for hot QCD studies starts with the first nuclear beams in 2010. The amount of information produced by this first run is overwhelming: The three experiments with nuclear program (ALICE, ATLAS and CMS) have provide new results in basically all subjects considered in previous experiments and have also shown the potential to make nuclear collisions at the TeV scale for the first time. I will review what the results from both RHIC and LHC imply for our understanding of hot and dense QCD matter from a theorists' perspective and how these new results change some of the concepts we developed in the last years.
Particular attention is devoted to the case of jets, as the first data recently published from the LHC and the limitations of previous approaches call for a new theory of jets in a medium.
\end{abstract}

\maketitle


\section{Introduction}

Hard probes are one of the pillars to characterize the hot and dense QCD matter created in high energy nuclear collisions. The presence of a large energy scale implies that a variety of distances are tested from the small distances of the very short times to the long distances at later times. An excellent example of such a hard probes is {\it jet quenching}, the modification of the jet structures while traveling a medium \cite{Bjorken:1982tu,Appel:1985dq,Blaizot:1986ma,Gyulassy:1990ye,Wang:1991xy}. A jet is the experimental messager of the original quarks and gluons created at the very early times by hard processes, those processes in which a large energy scale, the large transverse momentum of the created particles, is involved. In QCD the production cross section of a hadron $h$ with large transverse momentum, $p_T$, is computed  through the factorization formula which reads schematically
\begin{equation}
\frac{d\sigma^{AB\to h}}{dp_T}\sim f^A_i(x_1,Q^2)\otimes f^B_j(x_2,Q^2)\otimes\sigma^{ij\to k}(x_1,x_2,zp_T,Q^2)\otimes D_{k\to h}(z,Q^2)
\label{eq:fact}
\end{equation}
where along with the short distance parton-parton scattering cross section, $\sigma(x_1,x_2,zp_T,Q^2)$, long distance terms describing the structure of the colliding objects $A$ and $B$, the parton distribution functions (PDFs), $f_i(x,Q^2)$, and the hadronization of the quark or gluon $k$ into the final hadron $h$ appear. These two types of long-distance terms, which encode the non-perturbative contributions  from the large sizes (in QCD scales) of both the colliding and the produced hadrons, provide the tools to study QCD matter with hard processes. Indeed, the main goal of the present high energy nuclear collision programs both at RHIC and the LHC, as well as the previous SPS experiments, is to produce and characterize a new state of matter, the {\it Quark Gluon Plasma} (QGP),  in which some of the broken symmetries of the QCD Lagrangian are restored. In particular, at high enough temperatures and densities, a deconfined state of matter is expected \cite{Cabibbo:1975ig,Collins:1974ky,Freedman:1976ub,Shuryak:1978ij}. High energy nuclear collisions provide the experimental conditions to deposit a large amount of energy in a macroscopic region (again in QCD scales)  to allow for the needed collective behavior to develop. In this situation, one can imagine that if a hard process takes place {\it inside} this QGP, the long distance part of (\ref{eq:fact}) describing the hadronization of the original quark $k$ into a final hadron $h$, $D_{k\to h}(z)$,  will be modified and studying this modification leads to a better knowledge of the properties of the medium. This idealized picture implies, in particular, that the (possible) modifications of the other type of long-distance terms, the PDFs is under control and that the factorization formula (\ref{eq:fact}) applies.

This, in fact, defines the procedure which has been developed during the last 25 years since the first of the hard probes to QCD matter, the $J/\Psi$ suppression \cite{Matsui:1986dk}, has been proposed: (i) the {\it cold nuclear matter} background, which enters into a modification of the nuclear PDFs with respect to the proton PDFs or into a modification of the hadronization is normally constrained by control proton-nucleus experiments (deuteron-gold experiments were performed at RHIC for technical reasons) which fixes the benchmark on top of which {\it hot QCD matter} effects can be identified. (ii) on the other had, measurements of electroweak boson production are performed to check the normalization in nucleus-nucleus collisions. These measurements were performed for photons at RHIC and for either photons and Z/W bosons at the LHC. (iii) Hot matter effects are studied with different hard processes, including different quarkonia states ($J/\Psi$, $\Upsilon$, $\Upsilon'$...), high-$p_T$ particles (inclusive one- or two-particle distributions, reconstructed jets...), etc. A variety of probes  in two different experimental environments (RHIC and LHC) is available at present for an unprecedented characterization of the hot QCD matter. In the following I will review the present status.

\section{Initial state and {\it cold nuclear matter} background}

Hot QCD matter searches in high energy nuclear collisions involve background subtraction of those processes whose underlying dynamics is not related with the presence of a produced medium. These {\it cold nuclear matter effects} need to be under good control for benchmarking.
Assuming that the factorization (\ref{eq:fact}) holds, there are two main sources of background: the modification of the PDFs, $f^A(x,Q^2)$, in the nuclear environment when compared to the free proton PDFs and the modification of the hadronization $D_{k\to h}(z,Q^2)$ by the presence of nuclear {\it cold} matter. 

\begin{figure}
\begin{minipage}{0.5\textwidth}
  \includegraphics[width=0.8\textwidth]{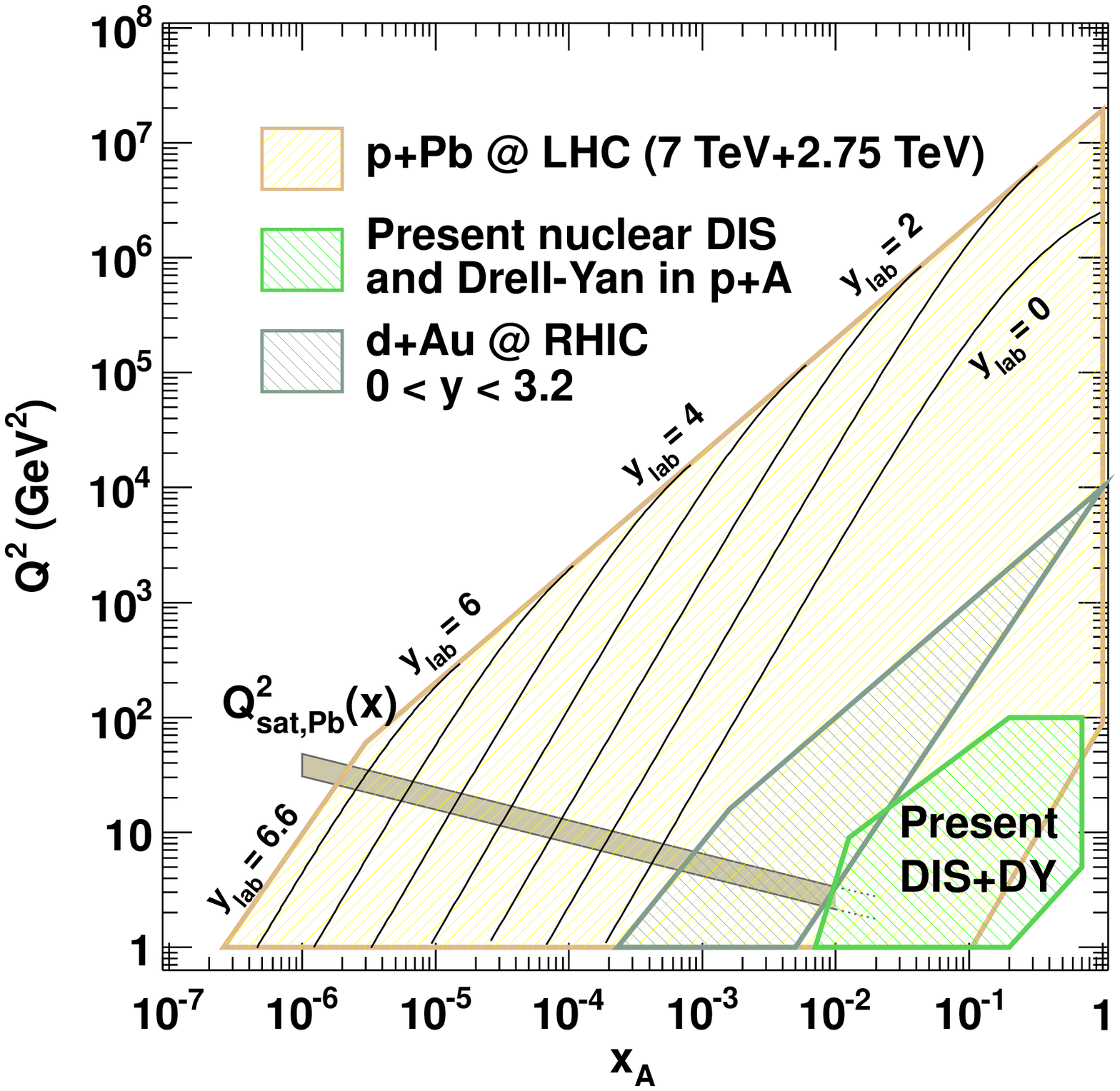}
  \end{minipage}
  \begin{minipage}{0.5\textwidth}
    \includegraphics[width=\textwidth]{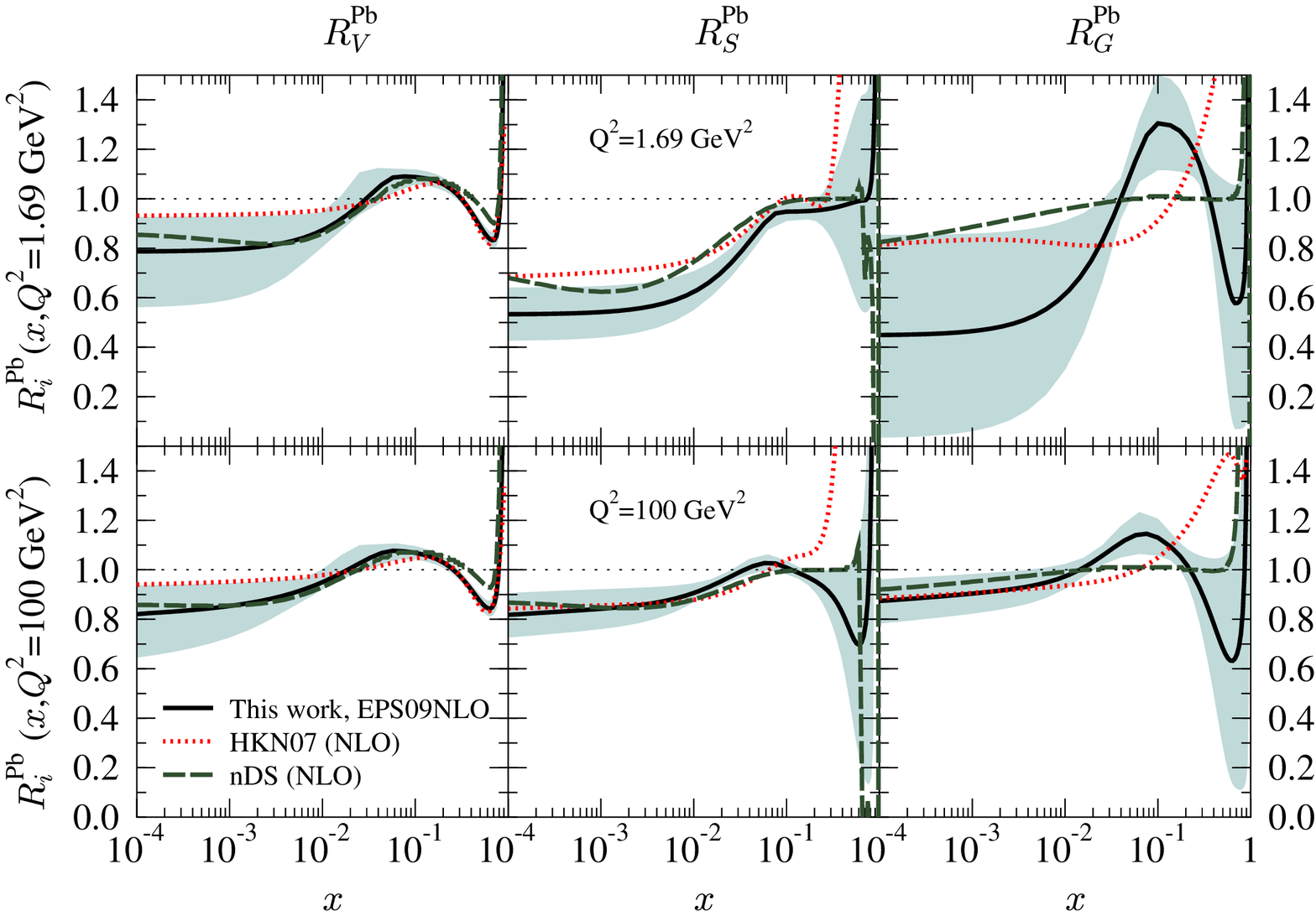}
  \end{minipage}
  \caption{(Left): Total kinematical reach of p+Pb collisions at $\sqrt{s}=$8.8 TeV  at the LHC for different rapidities in the laboratory frame. Also shown are the region of phase space studied by experiments of DIS with nuclei and Drell-Yan production in proton-nucleus collisions (the two main processes used in global nPDF fits) and the total reach of RHIC for $0<y<3$ --- Fig. from Refs.  \cite{Salgado:2011wc,Salgado:2011pf}. (Right): Nuclear parton distribution functions ratio with respect to the free proton case for two different virtualities and three different NLO analyses --- Fig. from Ref. \cite{Eskola:2009uj}.}
\label{fig:pA}
\end{figure}

The best way to constrain the nuclear PDFs is by experiments of lepton-nucleus deep inelastic scattering. The kinematical reach of the available data (which are, in fact, more than fifteen years old) is, however, rather limited, especially for the needs of the LHC program. In Figure \ref{fig:pA} the kinematical reach of the present  DIS and Drell-Yan (DY) data with nuclear targets (the data normally used to constrain nuclear PDFs) is plotted together with the needs for both RHIC and LHC kinematics \cite{Salgado:2011wc,Salgado:2011pf}. The central rapidities at RHIC were placed in the lucky situation of  overlap with previous DIS and DY data so that checks of the factorization equation (\ref{eq:fact}) and trustable subtraction of the background due to nuclear PDFs performed. The case of the forward rapidities at RHIC and basically the whole range of the LHC is, however, different. There is no experimental information which could constrain the nuclear effects in the small-$x$ region of interest for these two machines, which results into large uncertainties in the knowledge of the nuclear PDFs --- see Fig. \ref{fig:pA} (Right). In this situation, a parallel proton-nucleus collisions program will be needed at the LHC to reduce these uncertainties. It is worth mentioning that in the DGLAP approach, the nuclear effects to the PDFs rapidly disappear with increasing $Q^2$ in the small-$x$ region for the gluons and also for the quarks, although a bit slower. In the large-$x$, on the other hand, the uncertainties remain large in the whole range of $Q^2$. 

The error bands in the nuclear PDFs from Fig. \ref{fig:pA} translate into large uncertainties in the nuclear effects for some observables, especially those at the smaller virtualities. A recent experimental example reveals very clearly the need of constraining these uncertainties for a correct interpretation of the nucleus-nucleus data: ALICE Collaboration has measured the $J/\Psi$ suppression in Pb+Pb collisions in the forward rapidity region \cite{:2010px,Pillot:2011zg,MartinezGarcia:2011nf}. The suppression is about a factor of two when compared with the scaled p+p cross section and within the error bands of the suppression predicted by the EPS09 global fit of nPDFs presented in Fig. \ref{fig:pA}. Taken at face value, the suppression is then compatible with cold nuclear matter effects alone. Whether this is the case or not will need of better determination of the nPDFs which, at present, is only possible with p+Pb collisions at the LHC. Checks of the factorization hypothesis (\ref{eq:fact}) for the $J/\Psi$-production mechanism would also be interesting, in particular in view of the difficulties to theoretically describe this production cross section even in p+p collisions. 

As a final comment for this section let me mention that it is again a lucky situation that overlap exists between the RHIC and LHC kinematics whose full potential for benchmarking and nPDFs studies will be better exploited in proton-nucleus programs \cite{Salgado:2011wc,Salgado:2011pf}.

\section{Quarkonia suppression}

Quarkonia suppression is a conceptually simple and potentially powerful tool to characterize the properties of the produced QCD matter. An intuitive idea can be formulated in terms of the potential between a quark and an antiquark which, in the case of a hot deconfined medium is screened: if such a medium is created in a nuclear collision  bound states are disfavored during the plasma phase and the production of charmonia or bottomonia states suppressed  \cite{Matsui:1986dk}. The interpretation of the corresponding data has been confusing, however, in the last twenty years. The suppression has indeed been observed already in the pioneering experiments at the CERN-SPS of fixed-target S+U \cite{Abreu:1998wx} and Pb+Pb  \cite{Abreu:2000ni} or In+In collisions \cite{Arnaldi:2007zz}, it was also observed  at RHIC in Au+Au  \cite{Adare:2006ns,Reed:2011zz} and Cu+Cu collisions \cite{Adare:2008sh} and recently also at the LHC not only for the $J/\Psi$ \cite{:2010px,Pillot:2011zg,MartinezGarcia:2011nf} but also for excited quarkonia states \cite{Chatrchyan:2011pe}. One of the main problems for the interpretation of the data is the subtraction of the cold nuclear matter background. The suppression of both the $J/\Psi$, the $\Upsilon$ and other excited states has also been observed in proton-nucleus (or deuteron-gold) collisions in magnitude similar to the one in nucleus-nucleus collisions \cite{Alde:1990wa,Alde:1991sw,Abreu:1998ee,Adare:2007gn,Arnaldi:2010ky}. The theoretical description of this cold nuclear matter effects is not under good theoretical control and several mechanisms of $J/\Psi$-suppression are proposed. The most canonical one assumes a modification of the $J/\Psi$ yield due to nuclear PDFs and a modification of the hadronization modeled by a probabilistic Glauber model --- see e.g. \cite{Vogt:2004dh,Ferreiro:2008wc}. This factorization is not proved but used as a working hypothesis.

In this situation, the long-standing problem of the suppression of quarkonia states in nuclear collisions needs of a systematic study of the production in different systems (p+p, p+A, A+A) and energies as well as a systematic study of the different quarkonia states. Indeed, the excited states are predicted \cite{Digal:2001ue} to be more easily destroyed in hot matter than the ground states $J/\Psi$ and $\Upsilon$ --- a fact which is in qualitative agreement with the findings in \cite{Chatrchyan:2011pe} but whose quantitative understanding would need a better control over the cold nuclear matter effects. With the data accumulated in the last 20 years and the new data from both RHIC and, especially, the LHC a clear picture of this interesting observable should emerge in the near future.

\section{Jet quenching}

A quark or gluon produced at  high transverse momentum in an elementary QCD collision is associated with a large phase space available for extra gluon radiation. This extra radiation is emitted at small angles and can be experimentally identified in the form of jets. The theoretical control on the jet production and evolution is very good in the absence of a medium, this is, in fact, an essential requirement in the searches for new physics at the LHC. In a parton shower approach, the large virtuality of the original quark or gluon is reduced during evolution by radiating (mainly) gluons with a probability controlled by the Altarelli-Parisi splitting functions. The corresponding evolution equations of the fragmentation functions are known in different approximations. 

The case of the medium is not as well understood. Assuming that the evolution of the final state jet can be factorized from the initial state in a way similar to the vacuum, several different effect could appear: (1) {\it collisional energy loss}, due to elastic scatterings of the fast partons with the medium; (2) medium-induced gluon radiation also known as ({\it radiative energy loss}); (3) a modification of the color flow within the jet due to exchanges with the medium; (4) a modification of the ordering variable, or, in general, the evolution equations; etc..

RHIC phenomenology has been dominated by the energy loss mechanisms, this is because the corresponding jet quenching measurements were performed with inclusive particle measurements (one- or two-particle correlations) which measures the effects on the most energetic (leading) particle in the jet. A rather successful formalism (see e.g. Refs. \cite{CasalderreySolana:2007pr,d'Enterria:2009am,Wiedemann:2009sh,Majumder:2010qh} for recent reviews) based on the medium-induced gluon radiation is able to reproduce the corresponding data with two main unsolved issues 

\begin{enumerate}

\item {\it Heavy flavor suppression}: Basically all formalisms predict that heavy quarks will lose less energy than light quarks \cite{Dokshitzer:2001zm,Wicks:2005gt,Armesto:2003jh}. The exact difference depend on the details of the formalism but experimental data on the suppression of non-photonic electrons point to a stronger suppression \cite{Abelev:2006db,Adare:2010de}

\item {\it Sizable discrepancies between theoretical implementations}: The underlying physical hypothesis in the computations of the medium-induced gluon radiation are basically common to all formalisms but the actual approximations made translate into sizable differences in the output medium parameters \cite{Armesto:2011ht}.

\end{enumerate}

 \subsection{The theory for medium-induced gluon radiation}
 
 A convenient way of describing the propagation of very energetic particles traversing a piece of matter is in a semiclassical approach in which the medium is as a recoilless classical field (or equivalently a collection of static scattering centers if discretized). In the eikonal limit, when the energy of the quark is much larger than any other scale in the problem, the longitudinal and transverse degrees of freedom decouple --- here longitudinal refers to the direction along the propagation of the quark and one also has that $E\gg p_T$. In this setup, the propagation of a quark from a transverse position $\bf x$ at time $t_1$ to transverse position $\bf y$ at time $t_2$ is given by the Feynmann path-integral in transverse position
\begin{equation}
G({\bf x},t_1;{\bf y},t_2)=\int {\cal D}{\bf r}(t)\exp\left\{i\frac{E}{2}\int d\xi\left[\frac{d{\bf r}(\xi)}{d\xi}\right]^2+ig\int d\xi A({\bf r}(\xi))\right\}\, \longrightarrow\, W({\bf x})={\cal P}\exp\left\{ig\int d\xi A({\bf x},\xi)\right\},
\label{eq:pathint}
\end{equation}
which, in the asymptotic case, $E\to \infty$, recovers the Wilson line $W({\bf x})$. Here $A({\bf r}(\xi))$ is the color field of the medium at position ${\bf r}(\xi)$ and $\cal P$ refers to path ordering of the fields inside the integral. To continue, we notice that any observable will, in fact, involve colorless combinations of the Wilson lines, so, the simplest object which will appear in any calculation is $\frac{1}{N} {\rm Tr} \langle W({\bf x})W^\dagger({\bf y}) \rangle$, with $N=N_c$ for quarks and $N=N_c^2-1$ for gluons to average over the initial color configurations. The medium average $\langle...\rangle$ is a main ingredient in the calculation of the medium-induced gluon radiation spectrum. Several ways of performing these medium averages have been proposed. Multiple soft scatterings can be resummed in an exponential defining the transport coefficient $\hat q$
 \begin{equation}
\frac{1}{N} {\rm Tr}\langle W^{\cal y}({\bf x})W({\bf y})\rangle\simeq \exp\left\{-\frac{C_F}{2}\int d\xi n({\xi})\sigma({\bf y-x})\right\}
\simeq \exp\left\{-\frac{1}{4}\,\hat q_F\,L\,({\bf y-x})^2\right\}
\label{eq:multipleq}
\end{equation}
In the last term, the dipole cross-section has been written as $\sigma\simeq C\, {\bf r}^2$. This medium average is usually known as the BDMPS-Z approach, followed in Refs. \cite{Baier:1996kr,Zakharov:1997uu,Wiedemann:2000za} and also, to some extent, with a different treatment of the cross section $\sigma$ and the diffusion process, in AMY \cite{Arnold:2002ja}. For dilute media, the power-law tails of the potentials, neglected in (\ref{eq:multipleq}) may become relevant and an opacity expansion is usually employed. This is the medium average usually employed by the GLV group \cite{Gyulassy:2000er}, which, to some extent can be also identified in the twist expansion \cite{Wang:2001ifa}.

It is easy to check --- see e.g. \cite{CasalderreySolana:2007pr} for a derivation with the same notation used here --- that $\hat q$ is the average transverse momentum squared acquired by the system per mean free path in the medium,
$\hat q\simeq {\left\langle k_\perp^2\right\rangle}/{\lambda}$. This implies a rather general correspondence between energy loss and $k_T$ broadening $\Delta E\sim k_\perp^2 L/\alpha_s$.

The calculation of the medium-induced gluon radiation considers an fast (off-shell) quark or gluon produced in a hard process to compute a one-gluon inclusive spectrum. This setup returns no information about the multiple gluon emission and the corresponding medium-modification of the parton shower process. Most of the phenomenology of RHIC has been based on the independent gluon emission approximation \cite{Baier:2001yt,Salgado:2003gb,Wicks:2005gt}, the motivation being the medium-induced spectrum to be both collinear and infrared finite, in contrast to the vacuum one. Interestingly, a naive modification of the splitting function by the medium terms   in a DGLAP approach, results in very similar effects in the inclusive particle suppression \cite{Armesto:2007dt}.

\subsection{Jet quenching with inclusive particles}

\begin{figure}
 \begin{minipage}{0.5\textwidth}
  \includegraphics[width=0.9\textwidth]{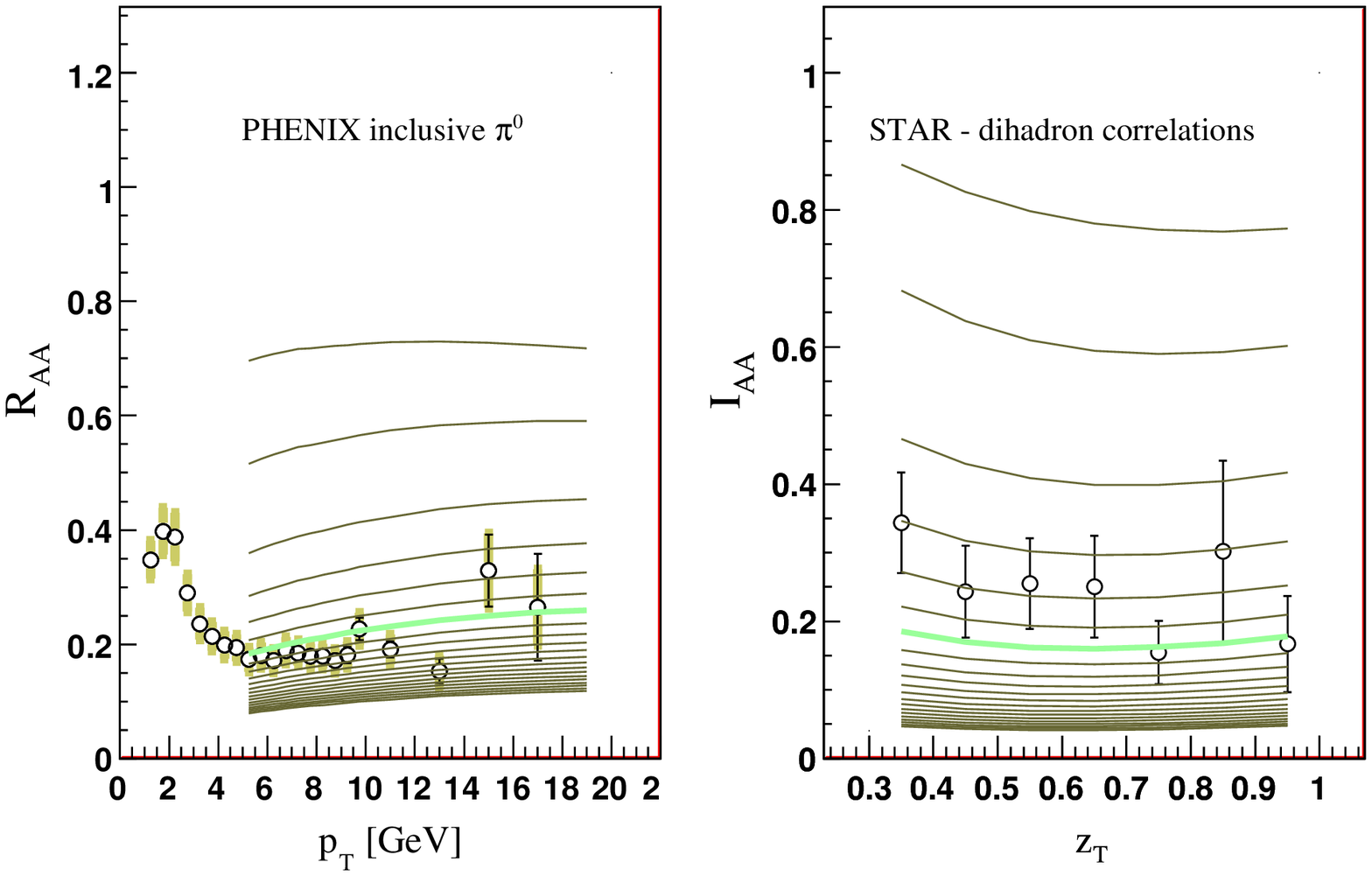}
   \end{minipage}
 \begin{minipage}{0.5\textwidth}
  \includegraphics[width=0.9\textwidth]{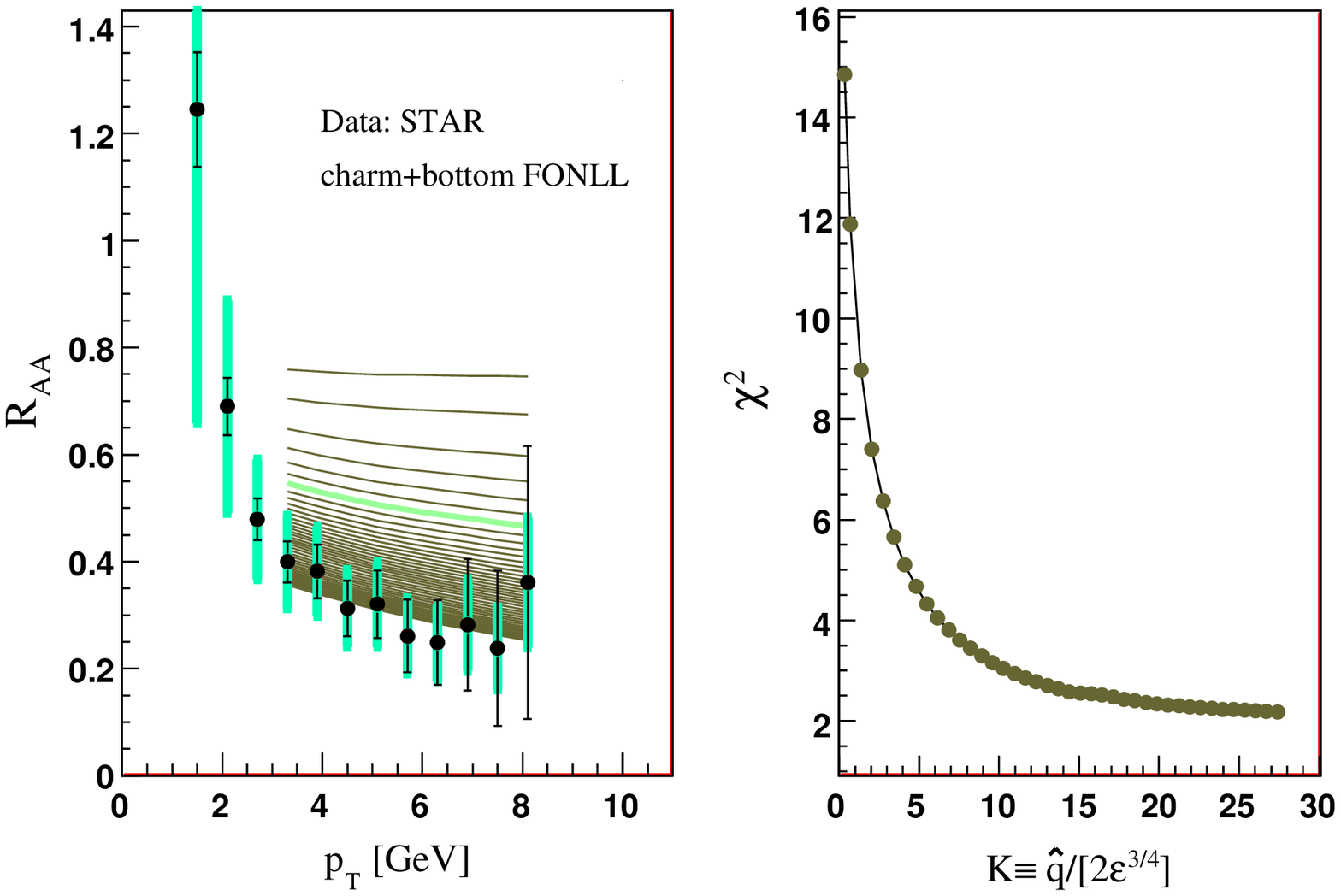}
\end{minipage}
  \caption{Nuclear modification factors $R_{AA}$ for single inclusive distributions  and $I_{AA}$ for double inclusive distributions for pions (first panel from the left) charged hadrons (second panel) and non-photonic electrons (third panel). The different lines correspond to $K$=0.5, 1, 2, .... 20. The last panel shows the values of $\chi^2$ computed for different $K$ for the case of the non-photonic electrons.  Fig. from Ref. \cite{Eskola:2009uj}.}
\label{fig:raa}
\end{figure}

The simplest observable of jets in nuclear collisions is the measurement of the one-particle inclusive production at high transverse momentum. The effect of the surrounding matter can be identified by the suppression of the signal, with respect to the proton-proton collisions, due to  energy loss. The nuclear modification ratio
\begin{equation}
R_{AA}=\frac{d\sigma^{AA}/dydk_T}{N_{\rm coll}d\sigma^{AA}/dydk_T}
\label{eq:raa}
\end{equation}
is normally employed to single-out the medium effects, where $N_{\rm coll}$ is a normalization factor computed in the Glauber model to allow the comparison with the proton-proton cross section. The suppression of high-$p_T$ hadrons is one of the first, and also one of the main, observations at RHIC \cite{Adcox:2004mh,Back:2004je,Arsene:2004fa,Adams:2005dq}. Several theoretical approaches have been used to reproduce the data, the most successful ones being those based on radiative energy loss as explained above. In Fig. \ref{fig:raa} we plot the description of the data in one of this approaches \cite{Armesto:2009zi} for both the one- and two-particle inclusive distributions (back-to-back signals for the second). 
The description of the data is  good. A quality analysis returns a value $K=4.1\pm0.6$ when the transport coefficient is parametrized as $\hat q= 2\,K\,\epsilon^{3/4}$, $\epsilon$ being the local energy density of the medium in a hydrodynamical approach. In the case of an ideal quark-gluon plasma, a free gas of quarks and gluons, $K\sim 1$ --- see e.g. \cite{Baier:2008js} --- indicating that the properties of the medium do not naively correspond to this simplified scenario. As mentioned above, however, despite the successful description of the data two main open issues need to be solved for which LHC data will be most helpful.

LHC collaborations have also measured $R_{AA}$ for inclusive particles at high-$p_T$ both for light hadrons \cite{Aamodt:2010jd,Lee:2011cs} and, interestingly, for charmed mesons \cite{Dainese:2011vb}. The  suppression for light hadrons turns out to be similar, though slightly larger, than the one at RHIC for moderate values of $p_T$. Models tested at RHIC can reproduce the data reasonably well, including the positive slope which indicates $R_{AA}\to 1$ for large transverse momenta. Concerning the $D$-meson suppression, with large error bars it also indicates a similar, although slightly smaller, suppression than the corresponding one for light hadrons. This was expected from calculations of medium-induced gluon radiation \cite{Armesto:2005iq}. So, there seems to be a compatibility of the well-tested approaches used in RHIC  phenomenology with the new data from the LHC. More quantitative analyses  should be performed now, with all available data, also when the medium density distributions from hydrodynamical analyses become available as input. 

\subsection{Reconstructed jets in nuclear collisions}

Although the results from the previous section are extremely interesting for the characterization of the medium properties, the use of inclusive quantities present also limitations which are difficult to overcome. In particular, in a scenario of very dense medium, surface effects could affect the extraction of the medium parameters and different approaches are difficult to distinguish. A powerful tool to overcome this limitations is the reconstruction of jets in nuclear collisions. In the ideal situation, if the whole energy of the jet can be reconstructed, the modifications that the medium induce in its structure gives a direct information about the splitting process as well as other mechanisms which could be present.

Jet reconstruction is one of the main issues in hadronic colliders, and essential for physics searches. In the case of the nuclear collisions, the size of the underlying event, with a very large multiplicity, makes the identification of the jets more difficult. The first data on identified jets has been performed at RHIC \cite{Putschke:2008wn,Salur:2008hs} and the first published data appeared very recently from the LHC \cite{Aad:2010bu,Chatrchyan:2011sx}. The analysis of the ATLAS and CMS collaborations present some surprising results. They can be summarized as follows:

\begin{enumerate}

\item Reconstructed jets from ATLAS are suppressed from central to peripheral collisions ($R_{CP}\sim 0.5$ and basically flat). This indicates that the sample of studied jets are still biased to some extent.
 
\item When the back-to-back jet signals are studied, the energy imbalance from the most energetic jet to the one in the opposite direction is  larger in central Pb+Pb  than in p+p collisions. This indicates a large energy loss of jets in the produced matter.

\item CMS data indicate that this energy lost is dominantly carried away by soft particles (less than ~2 GeV) at large angles. This contrasts with the vacuum where the particles are harder at large angles due to angular ordering.

\item The di-jet azimuthal asymmetry is very similar in Pb+Pb collisions and in p+p collisions. So, no strong change with respect to the vacuum jets is observed: the effect is not dominantly driven by e.g. emission of hard particles which would change the direction of the jets.

\item The fragmentation functions of the leading and the subleading jets do not present any change from Pb+Pb or p+p collisions. So, the fragmentation function is vacuum-like\footnote{Notice that the experimental fragmentation function (FF) is built by dividing the particles' transverse momentum by the jet's total energy, while the theoretical predictions before the data appeared typically dividing the transverse momentum of the particle by the energy {\it of the parton} originating the jet. In the second case (not possible in experimental conditions unless the whole energy of the jet is reconstructed) a suppression of the FF is predicted, in agreement with the suppression found in inclusive particle measurements as the ones in Fig. \ref{fig:raa}.}.

\end{enumerate}

Some of these properties were not, a priori, expected from theoretical estimates. In particular, the relation between broadening and energy loss mentioned above seems  difficult to reconcile, with those observations at least naively. However, a note of caution needs to be made in here as a complete picture of the underlying mechanism of jet quenching needs of a controlled analysis of several factors as, e.g. the amount of jets which are lost; the effect of the backgroud subtraction; the actual theoretical implementations which are compared with the data, etc. One can imagine, for example, a simplified scenario in which two different jet quenching mechanisms are at work and one of them is removed from the sample because it produces e.g. a too hard spectrum. With all these cautions, we can still try to extract some consequences, and  the properties above indicate that the effects in the measured jets are not compatible with a hard radiation at large angles, which would modify, in particular, the di-jet azimuthal asymmetry or with a strong modification of the radiation pattern inside the cone, which would modify the fragmentation functions. A naive interpretation of the data would then indicate that mechanisms in which the jet broadening and the energy loss do not follow the traditional relation $\Delta E\sim \langle k_T\rangle L/\alpha_s$ are favored in the particular sample of jets measured.

\subsection{Towards a new theory of jets in a medium}

The limitations in the theoretical implementation of jet quenching as well as the quality of the new data becoming available, especially from LHC, calls for a new theory of jets in a medium. An essential ingredient that any description of the jet development should contain is a correct treatment of the multi-parton emissions. The traditional way of assuming an independent gluon emission approximation is probably good enough to estimate the energy loss and, hence, for the phenomenology of inclusive particle suppression. The description of a  final state with a large number of gluons emitted needs, on the other hand, the inclusion of quantum interferences among different emitters which are known to be essential in the vacuum. As a first step towards this goal, recent developments consider the emission out of a quark-antiquark antenna \cite{MehtarTani:2010ma,MehtarTani:2011tz,MehtarTani:2011jw,CasalderreySolana:2011rz,Armesto:2011kh}. The setup captures the main physical ingredients in the vacuum, in particular, the presence of angular ordering due to color coherence effects. In the case of a medium the situation is radically changed. Several regimes have been identified, and interestingly, a new contribution emerges in which a vacuum-like radiation, but {\it antiangular ordered} \cite{MehtarTani:2010ma} can be identified. This new contribution is especially interesting because its features are completely different from all known medium-induced gluon radiation presented in the previous sections. This becomes more clear in the soft limit where the sum of the vacuum plus the medium-induced gluon radiation off a $q\bar q$ antenna with opening angle $\theta_{q\bar q}$ in a singlet state is simply
\begin{equation}
\label{eq:nqmed}
dN^{\rm tot}_{q,\gamma^*}=\frac{\alpha_s C_F }{\pi}\frac{d\omega}{\omega}\frac{\sin\theta \ d \theta}{1-\cos\theta}\left[\Theta(\cos\theta-\cos\theta_{q\bar q})-\Delta_{\rm{med}}\,\Theta(\cos\theta_{q\bar q}-\cos\theta)\right] \;.
\end{equation}
Here the first term is just the vacuum angular-ordered contribution and the second term is the new medium contribution which has been called {\it antiangular ordering} \cite{MehtarTani:2010ma}. In particular, and in contrast to previous results, a soft divergence appears also for the medium-induced part due to the vacuum-like spectrum. The parameter of the medium controlling the amount of {\it antiangular ordering} is the dipole scattering amplitude
\begin{equation}
\label{eq:Delta}
\Delta_{\rm med}=1-\frac{1}{N_c^2-1}\langle {\bf Tr}\,U_p(L,0) U_{\bar p}^\dag(L,0) \rangle \,,
\end{equation}  
which, by unitarity, is bounded by 1. In the case of an opaque medium, $\Delta_{\rm med}\to 1$, a {\it total decoherence} is then achieved in which the total spectrum is   \cite{MehtarTani:2011tz}
\begin{equation}
\label{eq:spectrumdecoh}
dN^{\rm tot}_{q,\gamma^*}\Big|_{\rm opaque} = \frac{\alpha_sC_F }{\pi}\frac{d\omega}{\omega}\frac{\sin\theta \ d \theta}{1-\cos\theta} = dN^{\rm tot}_{q,g^*}\Big|_{\rm opaque}
\end{equation}
The last equality means that another property of the spectrum is the {\it memory loss}: the radiated gluons do not keep information about the original pair being in a singlet or an octect state. Interestingly, these new properties survive the soft limit and the spectrum retains a form similar to (\ref{eq:nqmed}) for sizable values of the gluon energy.

These new results indicate that the medium-induced gluon radiation off a single emitter, considered up to now in all phenomenological approaches and also implemented in some Monte Carlo codes \cite{Lokhtin:2008xi,Zapp:2008gi,Armesto:2009fj,Armesto:2009ab,Schenke:2009gb}, would not be enough for a correct interpretation of the experimental data. Non-trivial structures appear when considering more than one emitter, the realistic situation in a jet shower, as already known from the vacuum. 

The features of the radiation are, on the other hand, in good qualitative agreement with the experimental data on jets presented in the previous sections: The spectrum (\ref{eq:nqmed}) presents vacuum-like radiation outside the cone delimited by the pair angle, in particular with a soft divergency, so, soft, vacuum-like, radiation is expected at relatively large angles while the radiation inside remains unchanged and just as in vacuum. These are qualitative behaviors which should be contrasted with data in a more quantitative analysis once the correct implementation of the multigluon emission is known.

\begin{theacknowledgments}
This work is supported by Ministerio de Ciencia e Innovacion of Spain, Xunta de Galicia, by project Consolider-Ingenio 2010 CPAN and Feder. CAS is a Ram\'on y Cajal researcher.
\end{theacknowledgments}



\bibliographystyle{aipproc}   

\bibliography{sample}

\IfFileExists{\jobname.bbl}{}
 {\typeout{}
  \typeout{******************************************}
  \typeout{** Please run "bibtex \jobname" to optain}
  \typeout{** the bibliography and then re-run LaTeX}
  \typeout{** twice to fix the references!}
  \typeout{******************************************}
  \typeout{}
 }

\end{document}